\documentclass[conference]{IEEEtran}
\IEEEoverridecommandlockouts
\usepackage{cite}
\usepackage{amsmath,amssymb,amsfonts}
\usepackage{algorithmic}
\usepackage{graphicx}
\usepackage{textcomp}
\usepackage{xcolor}
\usepackage{multirow}
\def\BibTeX{{\rm B\kern-.05em{\sc i\kern-.025em b}\kern-.08em
    T\kern-.1667em\lower.7ex\hbox{E}\kern-.125emX}}

\usepackage{booktabs}

\begin{document}

\title{
ORACLE: Collaboration of Data and Control Planes to Detect DDoS Attacks \\
}

\author{\IEEEauthorblockN{\textbf{Sebasti\'an G\'omez Mac\'ias}}
\IEEEauthorblockA{
Universidad de Antioquia\\
Medell\'in, Colombia \\
sebastian.gomez3@udea.edu.co}
\and
\IEEEauthorblockN{\textbf{Luciano Paschoal Gaspary}}
\IEEEauthorblockA{
Universidade F. do Rio Grande do Sul\\
Porto Alegre, Brasil \\
paschoal@inf.ufrgs.br}
\and
\IEEEauthorblockN{\textbf{Juan Felipe Botero}}
\IEEEauthorblockA{
Universidad de Antioquia\\
Medell\'in, Colombia \\
juanf.botero@udea.edu.co}
\and

}

\maketitle

\begin{abstract}

The possibility of programming the control and data planes, enabled by the Software-Defined Networking (SDN) paradigm, represents a fertile ground on top of which novel operation and management mechanisms can be fully explored, being Distributed Denial of Service (DDoS) attack detection based on machine learning techniques the focus of this work. To carry out the detection, this paper proposes ORACLE: cOllaboRation of dAta and Control pLanEs to detect DDoS attacks, an architecture that promotes the coordination of control and data planes to detect network attacks. As its first contribution, this architecture delegates to the data plane the extraction and processing of traffic information collected per flow. This is done in order to ease the calculation and classification of the feature set used in the attack detection, as the needed flow information is already processed when it arrives at the control plane. Besides, as the second contribution, this architecture breaks the limitations to calculate some features that are not possible to implement in a traditional OpenFlow-based environment. In the evaluation of ORACLE, we obtained up to 96\% of accuracy in the testing phase, using a K-Nearest Neighbor model.

\end{abstract}


\vspace{-0.2cm} 
\section{Introduction}
A DDoS attack is considered a powerful weapon against Internet security due to its ability to bring down even the largest websites by overloading servers' resources \cite{norton}. In turn, Software-Defined Networking (SDN) is an emerging network paradigm that decouples the control and data planes where the network control is logically centralized and programmable. Considering the different proposed DDoS detection approaches, several leverage the SDN advantages: global network view and control plane programmability. Among these mechanisms, those based on Machine Learning (ML) algorithms are considered the best in terms of key metrics such as detection speed, accuracy, and reliability \cite{ChicaJP,Sultana2019}.





A traditional ML-based detection system is mainly based on: i) a traffic information collecting mechanism, ii) a feature calculation module, and iii) a classification model; all of them implemented at the control plane. The collecting mechanism obtains traffic information from the packets that arrive at the SDN controller (\textit{Packet\_in}) and also from the statistics provided by OpenFlow (SDN southbound interface). The feature calculation module processes the collected information to build the set of descriptors, and, finally, the classification module classifies this set of features in search of possible ongoing DDoS attacks. However, in the implementation of these detection systems, we observe two design challenges, which we enumerate and describe next.

The first challenge is to prevent overloading the control plane due to the execution of tasks related to polling information from the network traffic, as well as the calculation and classification of the feature set. To carry out these tasks, the detection system has to store, update, and filter different information samples, which are sorted and processed per flow, connection, or source-destination pair. The execution of these tasks in the presence of large amounts of packets and flows can easily overload the control plane. 

The second challenge lies in extracting, in real-time, the information needed to calculate the feature set selected as the best traffic descriptors. In an OpenFlow-based SDN environment,  limited processing can be performed on a per-packet basis at the data plane. The extraction of sufficient traffic information with fine granularity needed to calculate a wide variety of features in real-time becomes impractical. Therefore, from a huge amount of traffic features proposed in the literature, just a specific group of them (e.g., flow's byte/packet count and duration) are possible to implement.




In order to overcome these challenges, we propose ORACLE, an approach based on the cOllaboRation of dAta and Control pLanEs to detect DDoS attacks using the paradigm of programmable data planes. The proposed approach fully explores the programming of packet processing at a switch level. To program the data plane, we use the P4 programming language \cite{p4} together with the P4Runtime interface (to provide communication with the control plane). At the data plane, we propose a hash-based data structure to store customized per-flow information. We also propose a mechanism to periodically report such information to the control plane, where the feature set is subsequently calculated and classified per each reported flow. The main contributions of ORACLE are:





\begin{itemize}
    \item A novel strategy that is expected to be deployed over the proposed SDN-P4 architecture to periodically calculate specific flow features to be used by the detection system. 
    \item An architecture that delegates to the data plane the tasks of collecting, processing, and storing the required information per traffic flow. This facilitates the calculation and classification of the feature set by the control plane. The architecture is not limited only to DDoS attacks. On the contrary, it provides the framework for the calculation of any feature needed to detect other attacks.
\end{itemize}

\section{Background and Related Work}

According to \cite{typeflows}, for traffic classification, the list of potential features is divided in five categories: \textit{packet} level (e.g., packet length mean), \textit{flow} level (e.g., average packet length and packets per flow), \textit{connection} level (e.g., advertised window sizes), \textit{intra-flow} level (e.g., features based on inter-arrival times between packets of the same flow), and \textit{multiflow}. 

In order to extract information at the flow or intra-flow level, three flow definitions apply \cite{typeflows}: \textit{unidirectional} flow, made up of packets sharing the same fields (source/destination IP, source/destination TCP/UDP ports and transport protocol number); \textit{bidirectional} flow, made up of a pair of unidirectional flows that are going in opposite directions between the same source and destination IP addresses and ports; and \textit{full} flow, a bidirectional flow captured during its entire lifetime.

Recent research efforts indicate that SDN is suitable for the implementation of sophisticated software solutions that allow detecting DDoS attacks. The following proposals, implemented at the control plane, make use of different feature sets and ML-based detection techniques.


In \cite{IntrusionDS}, the authors perform the DDoS detection by polling, each second, OpenFlow statistics per each unidirectional flow. To calculate the feature set, they store several samples of the same bidirectional flow (computed in the control plane from the unidirectional flows) to calculate features based on average and standard deviation (std) at flow level. Results show 96\% of accuracy with the Random Forest (RF) classifier.


The authors of \cite{atlantic} propose an integrated framework that comprises two operational phases at the control plane: 1) a lightweight processing phase that consists of monitoring the flows looking for an entropy anomaly, and 2) a heavyweight processing phase that is only invoked when an anomalous flow is detected.  The heavyweight process uses OpenFlow statistics (packet\_count, byte\_count, duration) at flow level to detect DDoS attacks. A Support Vector Machine (SVM) is used as the ML-classifier providing a detection accuracy score of 88\%. This means that the reduction of heavy processes at the control plane is achieved at big costs in terms of accuracy.


In the previous detection systems, and also in others such as \cite{yoon2015enabling}, we noticed that the implemented features are limited to a specific group coming from header field values extracted from the OpenFlow \textit{packet\_in} event. In \cite{CICIDS2017}, we can find a wide variety of features that one could implement to detect different attacks (e.g., DDoS). Among the huge list of features, those at the intra-flow level and several at the connection level are not possible to calculate in an SDN/OpenFlow environment. Consequently, we propose ORACLE, where the calculation of such features is possible, in real-time, via the programming and coordination of the data and control planes.

\section{ORACLE: cOllaboRation of dAta and Control pLanEs to detect DDoS attacks}

\subsection{Architecture}

\begin{figure}[]
    \centering
    \includegraphics[scale=0.44]{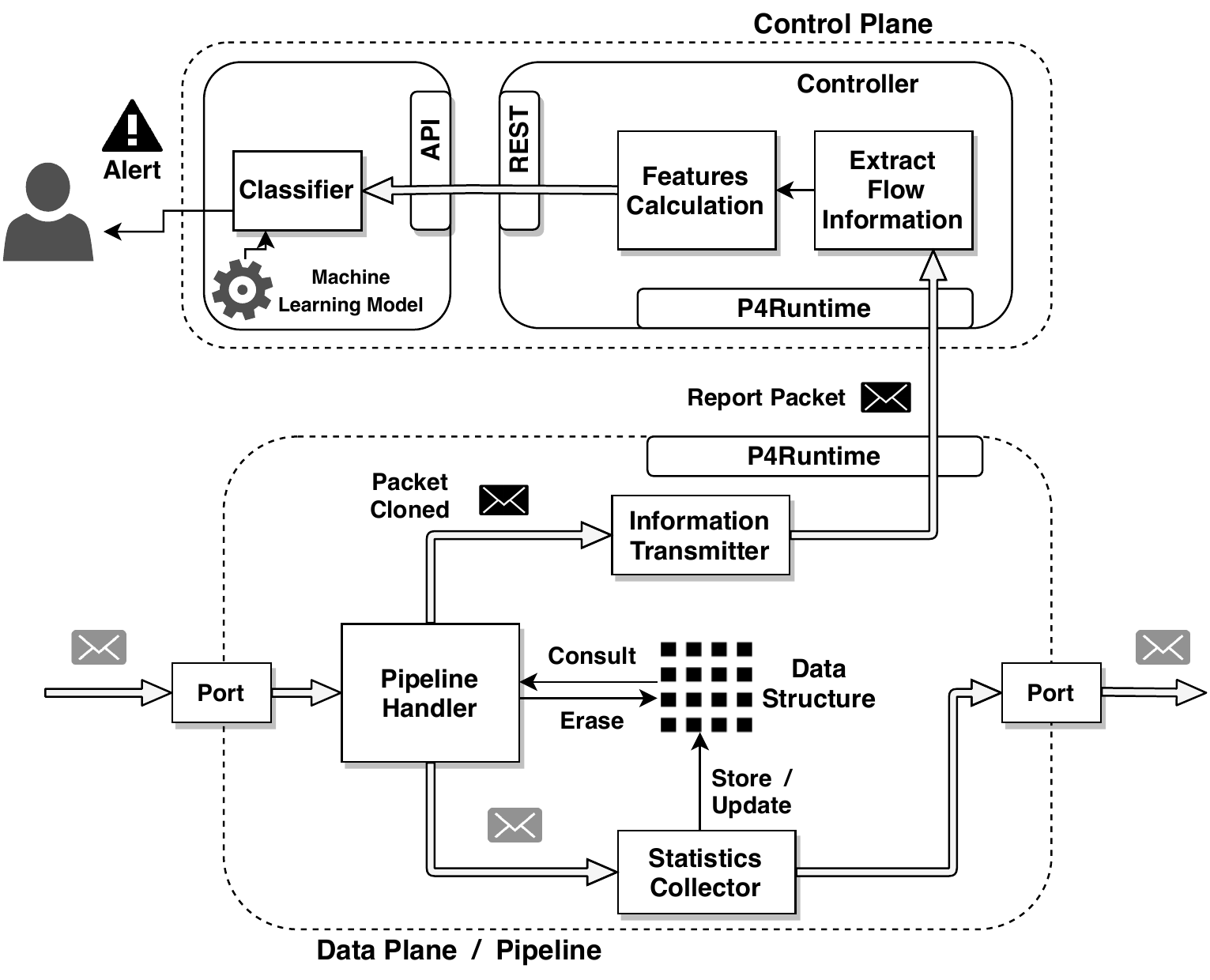}
    \caption{ORACLE system architecture}
    \label{fig:Arch}
\end{figure}

ORACLE is organized in two modules, namely \textit{Data Plane} and \textit{Control Plane}, which we show in Fig. \ref{fig:Arch} and describe next.


\textit{Data Plane}: ORACLE relies on a time window-based implementation. During each window, the data plane performs two tasks simultaneously: it i) collects new flow information, and ii) reports flow information collected and processed in the previous time window to the control plane. The data plane's module is composed of the following four components, developed using P4 primitives such as registers, clone packet, hash functions, and basic binary operations (+,-,*).

\begin{itemize}
    \item \textit{Data Structure}: It is used for the statistics organization, management, and storage per unidirectional flow. To access them, the flow identifier (flow-id) is mapped, through a hash function, to the respective index (position where the statistics are stored) in the data structure.

    \item \textit{Statistics Collector}: In charge of extracting and processing the needed information from the bit string of every packet traversing the device. With this information, the statistics of the flow associated with the packet are updated.

    \item \textit{Pipeline Handler}: In charge of controlling the time window duration and validating if the data structure contains new flow information to report. If so, the incoming packet is cloned and forwarded to the Information Transmitter component along with the per-flow information that will be subsequently sent to the controller.
    
    \item \textit{Information Transmitter}: This component transforms the cloned packet in a report packet that is forwarded to the control plane using the P4Runtime interface. The cloned packet is modified by eliminating its payload and adding the flow information as a new custom header.
    
\end{itemize}



\textit{Control Plane}: It is composed of an SDN controller and a traffic classification component connected via REST-API. To classify the reported flows, three components are involved:

\begin{itemize}
    \item \textit{Extract Flow Information}: This component extracts the custom header from the report packet and parses it to get the information of every flow sent by the data plane. 
    
    \item \textit{Feature Calculation}: It builds a tuple with the features set calculated with the flow information.
    
    \item \textit{Classifier}: An already loaded ML model classifies every tuple received in DDoS or Benign class. When a possible attack is detected, an alarm is issued.
\end{itemize}


\vspace{-0.1cm}
\subsection{Data Plane Storage Mechanism}

\begin{figure}[]
    \centering
    \includegraphics[scale=0.38]{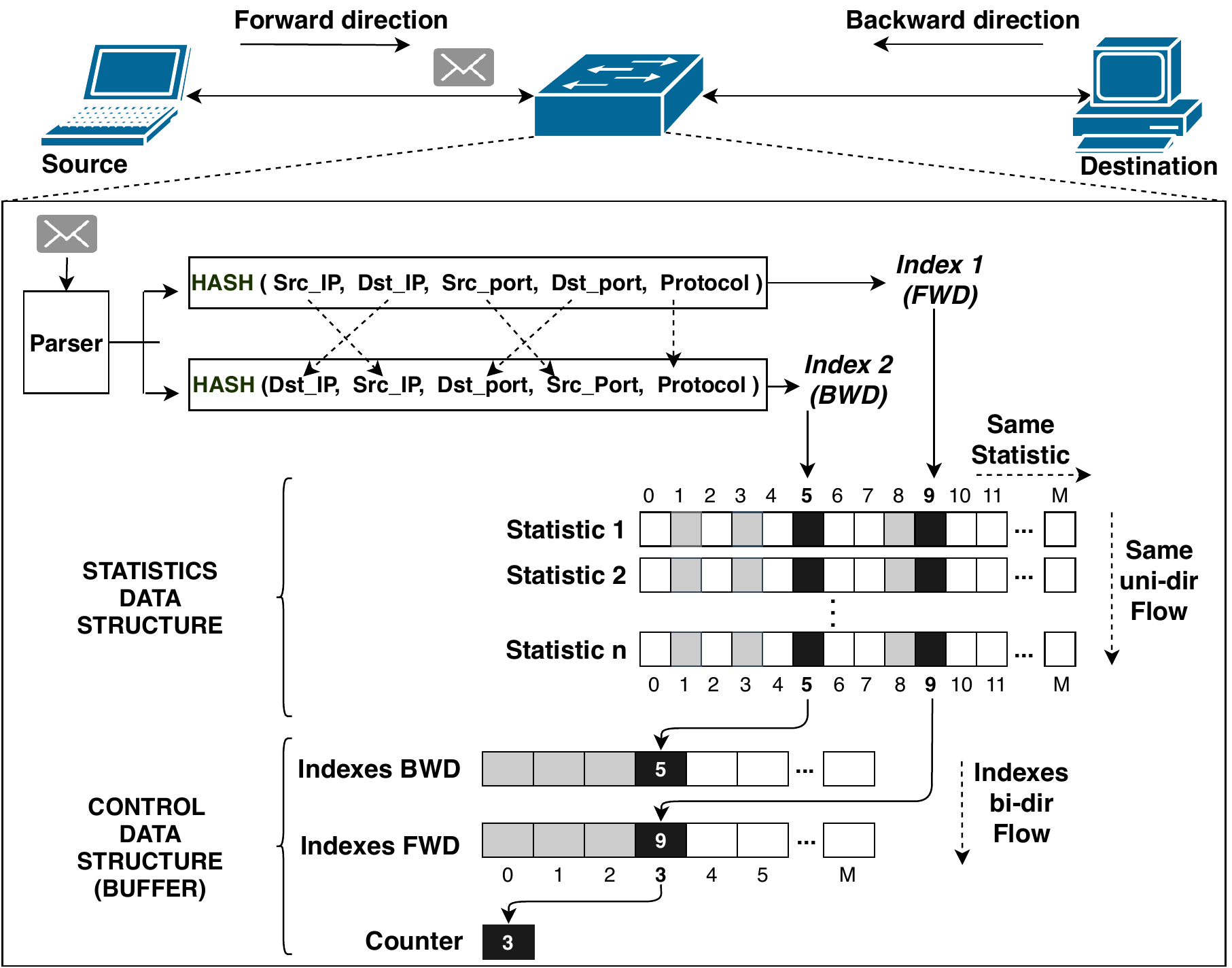}
    \caption{Mechanism to store statistics per flow.}
    \label{fig:GetStatisticsModule}
\end{figure}

ORACLE collects and stores statistics of packets belonging to each flow on the programmable switches. It guarantees that flow statistics can be consulted and updated each time a new packet belonging to the flow arrives to the P4 device.

We use P4 registers to define a data structure to store flow statistics. A register is a fixed-size array that uses an index to point to the stored elements. To consolidate the structure, we use a set of such registers to build a matrix-like structure, where each row stores values of the same statistics belonging to different flows, and each column stores different statistics of the same unidirectional flow (see Fig. \ref{fig:GetStatisticsModule}). To designate a unique index (column) to each unidirectional flow, we use the hash function with an input value (Flow-Id) that consists of source/destination IP, source/destination port, and transport protocol. The hash's output is the index mapped to the flow.

Figure \ref{fig:GetStatisticsModule} shows how per-flow statistics are stored and updated by the Statistics Collector component. Once a packet enters the P4 device, its bit string is parsed to extract and map each header field value to P4 variables. From these variables, the collector builds the 5-tuple (Flow-Id) to calculate the hash function. Once having the flow index, the flow statistics are updated in the data structure with the needed information that is obtained by processing the P4 variables.


As we are working with bidirectional flows and the method stores unidirectional flows, it is important to identify the pair of flows (indexes) defining the bidirectional flow within the data structure. To know the index of the opposite direction of the flow, we calculate again the hash function exchanging the order of the flow identifier fields (see Fig. \ref{fig:GetStatisticsModule}). Hence, we define forward (FWD) as the direction of the first packet of the flow and backward (BWD), the opposite direction.

\vspace{-0.1cm}
\subsection{Data Plane Flow Reporting Mechanism}

The Pipeline Handler component is in charge of reporting flow information to the control plane at the beginning of each time window. In order to forward the information of all bi-directional flows collected at the previous time window to the control plane, it is necessary to know the indexes that point to the specific flows within the data structure. To ease this process, each time a new flow is instantiated at the statistics data structure, both indexes (FWD and BWD) are stored in a new data structure composed of two registers and a counter (see Fig. 2). One register stores the indexes of the FWD unidirectional flows, and the other stores the indexes of the BWD flows. Hence, each column will contain the indexes of a different bidirectional flow and the counter will be just a reference to know the position of the last pair of stored indexes. This data structure works like a LIFO buffer: indexes are stored sequentially as the flows are instantiated. When the information is to be sent to the control plane, the indexes are consumed from the buffer in the reverse arrival order.

\begin{figure}[]
    \centering
    \includegraphics[scale=0.45]{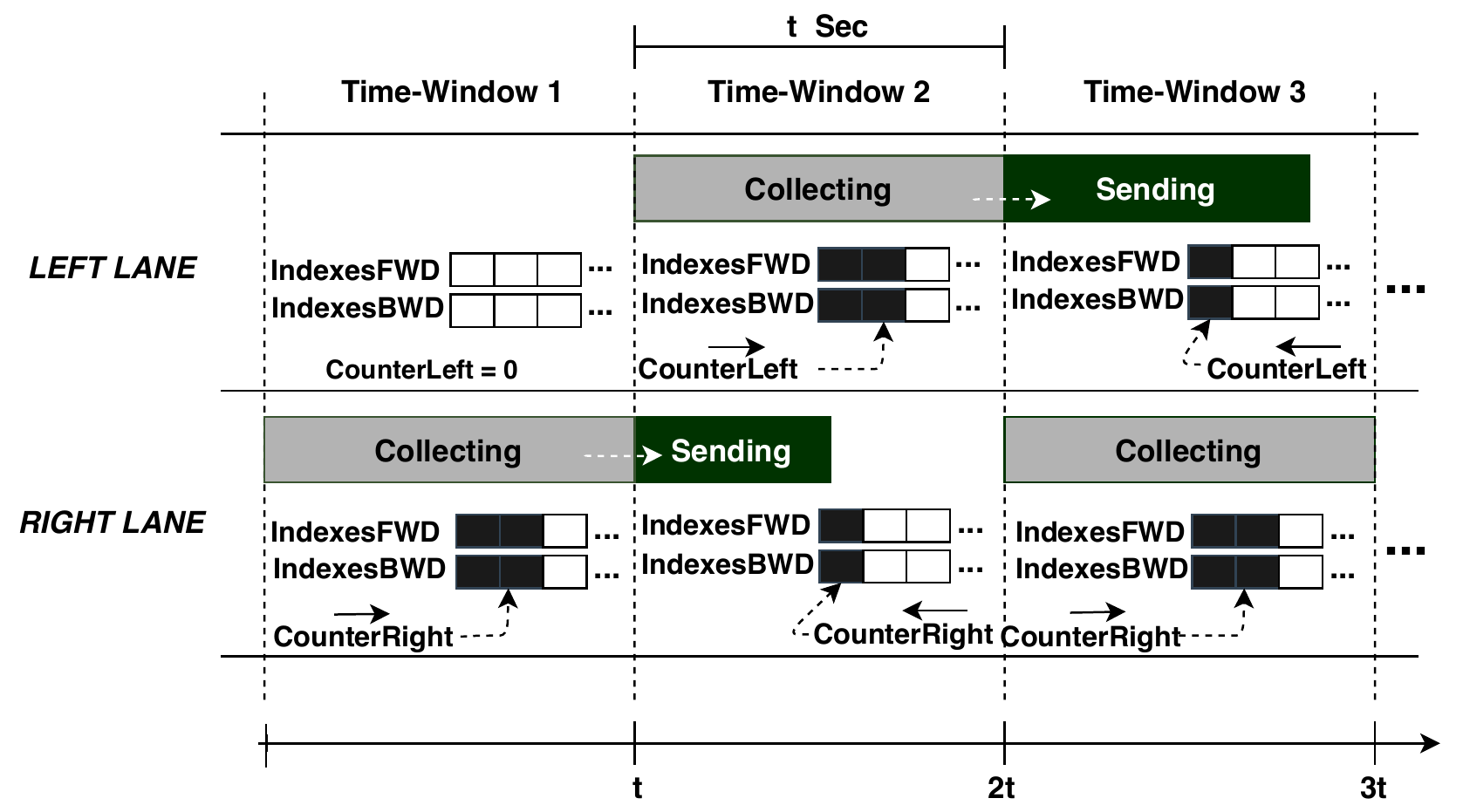}
    \caption{Lane-based mechanism to handle time windows and buffers.}
    \label{fig:handler}
\end{figure}

The implementation of the control data structure with a single buffer implies that the data plane interrupts the process of collecting information from new flows until the buffer is emptied. As the system must be able to perform both tasks (collecting and reporting flows to the control plane) simultaneously, we propose to abstract the idea of two independent monitoring ``lanes'', each one with its buffer. Therefore, is possible to delegate, over a same time window, one task to each lane, avoiding the process to be interrupted (see Fig. \ref{fig:handler}). 

Using this strategy based on independent buffers to store the indexes of bidirectional flows, the Pipeline Handler component can validate, in each time window, if there are flow statistics available to forward to the control plane. If so, this component looks for per-flow statistics from the data structure, clones the number of incoming packets needed to transmit this information, and delegates the transmitting process to the respective component. The Information Transmitter component embeds the information of each flow into a custom header. The number of fields (flows) within this header should be as big as the network MTU value to avoid injecting too many report packets into the network. 


\section{Detection System Implemented on ORACLE}

As proof of concept, we implemented a DDoS attack detection system that uses the feature set listed in Table \ref{tabla:features}. These features are recommended in \cite{CICIDS2017}, where the authors used the RandomForestRegressor technique to select, among 80 features, a small set considered the most adequate for DDoS attack detection. This set is made up of two features at flow level (F1 and F3) and two at intra-flow level (F2 and F4). On the other hand, there are features that are calculated considering the statistics in both flow directions (bidirectional), and one of them (F4) considering only the BWD direction.



\begin{table}[]
  \begin{center}
    \caption{Selected feature set}
    \label{tabla:features}
    \resizebox{8.7cm}{!}{
    \begin{tabular}{c|l|l|c}
      \toprule 
      \textbf{\#} & \textbf{Feature name} & \textbf{Description} & \textbf{Level / Type} \\
      \midrule 
                     F1 & FlowDuration & Flow duration in microseconds & Flow / \\ &&& (Bidirectional) \\
                    \hline  
                    F2 & FlowIATStd & Standard deviation of packet inter-arrival & Intra-Flow / \\ &&  times of the same flow. & (Bidirectional) \\ 
                    \hline
                    F3 & avgPacketSize & Average of packet payload sizes of the & Flow / \\ &&  same flow. & (Bidirectional) \\
                    \hline
                    F4 & BwdPktLenStd & Standard deviation of packet payload & Intra-Flow / \\ && sizes of the same flow in the backward & (Unidirectional) \\ &&  direction & \\
      \bottomrule 
    \end{tabular}}
  \end{center}
\end{table}

In Table \ref{tabla:features}, F2, F3, and F4 are based on average (avg) and standard deviations (std) operations, calculated using Eq. \ref{eq:mean} and \ref{eq:std}, respectively. $x_{i}$  corresponds to an information unit extracted from packet $i$, which belongs to a group of $n$ packets (of a same flow). The meaning of $x_{i}$ changes depending on the feature to calculate; for F2, it corresponds to the packet inter-arrival time (IAT), and for F3-F4, it corresponds to the size in bytes of the packet's payload. 

\begin{equation}\label{eq:mean}
    \footnotesize
    avg = \overline{x} = \frac{\sum_{i:1}^{n} x_{i}}{n - 1}
\end{equation}
\vspace{-0.22cm}
\begin{equation}\label{eq:std}
    \footnotesize
    std = \sqrt{\frac{\sum_{i:1}^{n}\left ( x_{i} - \overline{x} \right )^{2}}{n-1}} = \sqrt{\frac{\sum_{i:1}^{n}x_{i}^{2}-2\overline{x}\sum_{i:1}^{n}x_{i}+n\overline{x}^{2}}{n-1}}
\end{equation}

In an SDN/OpenFlow environment, the features at flow level (F1 and F3) can be calculated from OpenFlow statistics. The real challenge is to calculate the intra-flow features (F2 and F4) because: i) OpenFlow does not permit to know the packet IAT, and ii) to calculate the \textit{std} of an intra-flow measure, every $x_{i}$ value is required at the control plane to subtract the mean from it when the sum is being computed (see Eq. \ref{eq:std}, left side).

ORACLE allows overcoming these limitations as it stores individual information for each packet, even IAT values, and reports it periodically to the control plane. However, this may result in high latency and storage/processing resource consumption. A more ambitious solution would be to calculate these features directly on the data plane and avoid sending every $x_{i}$. This solution is not possible because programmable data planes are usually limited in the available mathematical operations (e.g., division and square root are not allowed).

\begin{table}[]
  \begin{center}
    \caption{Information per-Flow Reported by the Data Plane}
    \label{tabla:statistics}
    \resizebox{8.2cm}{!}{
    \begin{tabular}{c|l|l}
      \toprule 
      \textbf{\#} & \textbf{Feature / Statistic} & \textbf{Description} 
      \\
      \midrule 
                    F1 & FlowDuration & Flow duration in microseconds  \\
                    \hline
                    S2 & TotFwdPkt & Total number of packets in the FWD direction  
                    \\
                    \hline
                    S3 & TotBwdPkt & Total number of packets in the BWD direction   \\  
                    \hline
                    S4 & TotLenFwdPkt & Aggregation of packets' payload size in FWD \\ && direction $\left(\sum x_{i}\right)$  \\
                    \hline
                    S5 & TotLenBwdPkt & Aggregation of packets' payload size in BWD \\ && direction $\left(\sum x_{i}\right)$ \\
                    \hline
                    S6 & TotLenBwdPktSqrt & Aggregation of packets' payload size  \\ && square in BWD direction $\left(\sum x_{i}^{2}\right )$  \\
                    \hline
                    S7 & IATTotal & Aggregation of the packet inter-arrival times \\ && $\left(\sum x_{i}\right)$ \\
                    \hline
                    S8 & IATTotalSqrt & Aggregation of the packet inter-arrival times \\ && square $\left(\sum x_{i}^{2}\right )$ \\
                    
      \bottomrule 
    \end{tabular}}
  \end{center}
\end{table}

Therefore, we propose the following alternative. The traditional \textit{std} (Eq. \ref{eq:std}) is transformed using arithmetic properties to have the right side representation. Now, this new representation depends on two sums: the aggregation of values $\left(\sum x_{i}\right )$, and the aggregation of square values $\left(\sum x_{i}^{2}\right )$. Using the data plane mechanisms proposed in ORACLE, we calculate, directly at the data plane, the result of both sums per each flow, as the on-going packets traverse the data plane device. Therefore, when these aggregation results are reported together with the packet counters to the control plane (see Table~\ref{tabla:statistics}), the controller has only to replace them in Eq. \ref{eq:mean} and \ref{eq:std} to calculate F2, F3, and F4 per each flow in an easy and fast way, as specified below. Regarding F1, it is calculated directly in the data plane and reported to the control plane together with the statistics.


\begin{itemize}
    \item \textit{FlowIATStd (F2):}
        \begin{equation}\label{eq:nF2}
            \footnotesize
            n = S2 + S3 \quad (Bidir.) \quad\quad  ; \quad\quad \overline{x} = \frac{S7}{n - 1}
        \end{equation}
        \vspace{-0.25cm}
        \begin{equation}\label{eq:stdF2}
            \footnotesize
            std = \sqrt{\frac{S8-2*\overline{x}*S7+n\overline{x}^{2}}{n-1}}
        \end{equation}
        
    \vspace{-0.25cm}
    \item \textit{avgPacketSize (F3):}
        \begin{equation}\label{eq:nF3}
            \footnotesize
            n = S2 + S3 \quad (Bidir.) \quad\quad  ; \quad\quad \overline{x} = \frac{S4 + S5}{n - 1}
        \end{equation}
        
    \vspace{-0.25cm}    
    \item \textit{BwdPktLenStd (F4):}
        \begin{equation}\label{eq:nF2}
            \footnotesize
            n = S3 \quad (Unidir.)  \quad\quad  ; \quad\quad  \overline{x} = \frac{S5}{n - 1}
        \end{equation}
        \vspace{-0.25cm}
        \begin{equation}\label{eq:stdF3}
            \footnotesize
            std = \sqrt{\frac{S6-2*\overline{x}*S5+n\overline{x}^{2}}{n-1}}
        \end{equation}

\end{itemize}

\vspace{1mm}

\vspace{-0.15cm}
\section{Evaluation}
\subsection{Experimental Setup and Methodology}

We replicated the packet trace used in \cite{CICIDS2017} over a real scenario. This experiment consisted of a 20 minute-long DDoS attack generated from different devices located in an external network, trying to deny the service provided by a web service located in the DMZ of a local network. This trace, which has traffic captured for 90 minutes, is composed of 762,973 DDoS packets and 1,382,900 benign packets, stored in a PCAP file.

\begin{figure}[] 
    \centering
    \includegraphics[scale=0.4]{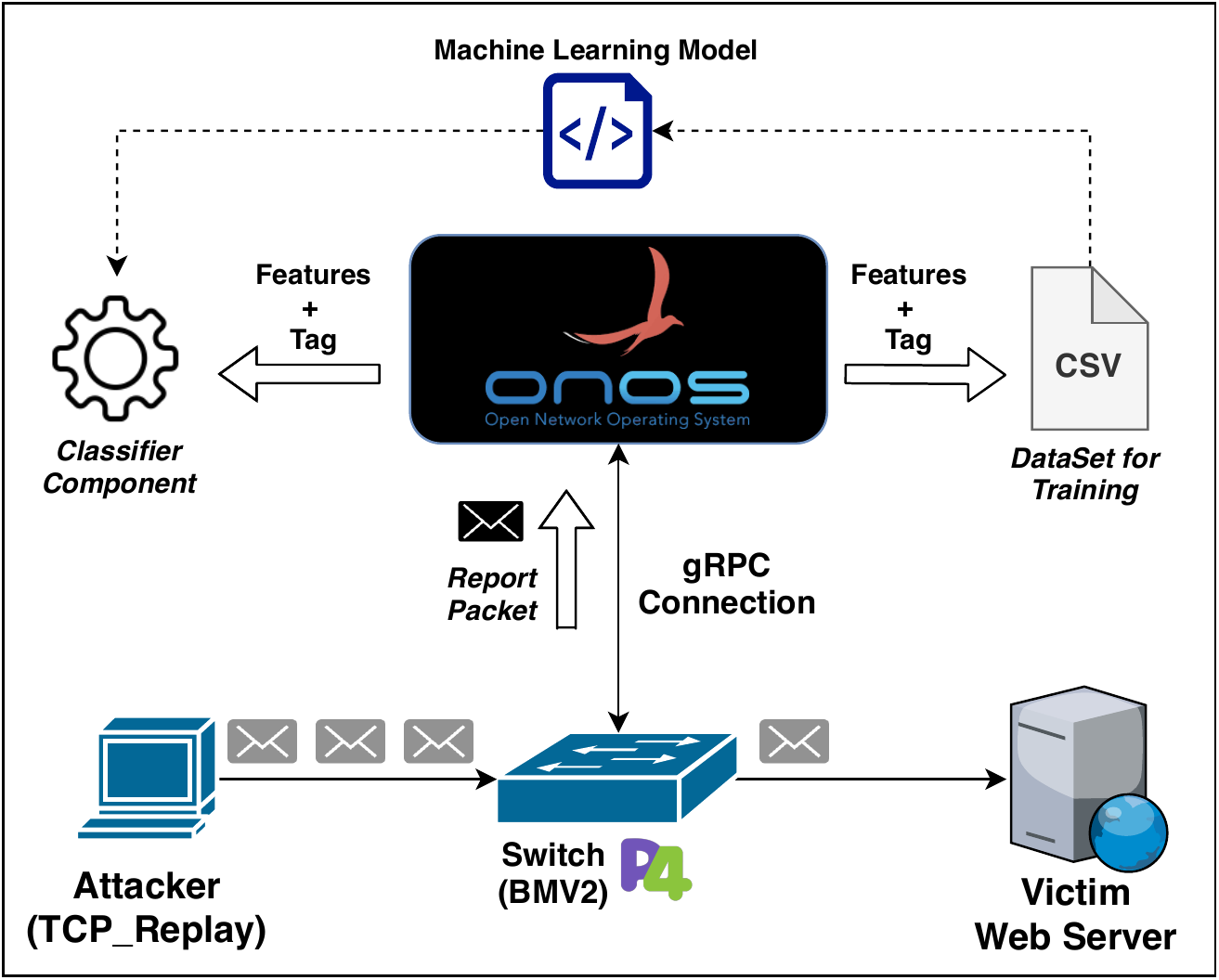}
    \caption{Experimental setup.}
    \label{fig:scenario}
\end{figure}

Fig.~\ref{fig:scenario} shows the implemented scenario. The data plane, emulated using Mininet, consists of: i) a BMV2 virtual switch with P4 support (in charge of executing the data plane strategy);  ii) a host responsible for reproducing the traffic trace using the TCPReplay tool (representing the external network where the DDoS attack is originated); and iii) a victim host (Web Server), where all replicated traffic is forwarded to. In turn, the control plane is composed by the ONOS controller, which is connected to the BMV2 switch using the P4Runtime interface. This plane has a classifier component, which loads a previously trained ML-model. This scenario was implemented on a Core i5 5200 computer, with 12 GB of RAM and the Ubuntu 18.04 LTE operating system.



The implementation of our proposed system was exposed to two different workloads, one for the training phase of the ML model (WL1) and another for the testing phase of the same model (WL2). To create both workloads, the pcap file was divided into 18 sets of 5 minutes each, labeled from 1 to 18 considering their order in the trace. The sets with odd numbering were put together to build WL1. Then, the sets with even numbering were also joined to build WL2. 

Before training the classification model, we had to build a database containing flow information. To create it, an ONOS application was developed to store, inside a .csv file, the feature set, and the respective tag (class: ``Benign", or ``DDoS") for every flow reported by the data plane (see Fig \ref{fig:scenario}). These tags, needed to train the supervised model, were obtained directly from the packets belonging to WL1. They were previously marked (00: Benign, 11: DDoS) by modifying the last two bits of the IPv4 ToS header field.

Besides focusing on training the best ML model to detect DDoS attacks, we also decided to investigate how the detection system could be affected by different time window values at the data plane. For this assessment, we built a different training dataset per each time window value. Table \ref{tabla:datasets} shows the characterization of the training datasets constructed considering different time window values (5, 20, 40, and 60 seconds).

\begin{table}[h]
  \begin{center}
    \caption{Training Dataset Characterization}
    \label{tabla:datasets}
    \resizebox{8cm}{!}{
    \begin{tabular}{c|c|c|c|c}
      \toprule 
      \multicolumn{5}{c}{\textbf{Dataset per Time Window Duration}} \\
      \midrule
      \textbf{Tag / Time Window } & \textbf{5 Sec} & \textbf{20 Sec} & \textbf{40 Sec} & \textbf{60 Sec} \\
      \midrule 
            Benign Flows & 61,736 & 40,462 & 31,894 & 29,670 \\
            DDoS Flows & 49,012 & 45,688 & 44,511 & 43,255  \\
            \midrule
      \textbf{Total Flows per Dataset} & 110,748 & 86,150 & 76,405 & 72,925 \\
      \bottomrule
    \end{tabular}}
  \end{center}
\end{table}

After creating the mentioned databases, the next step consisted of training the ML models in an offline manner. We used and compared two ML classifiers per each time window: RF and K-Nearest Neighbor (KNN) (8 models in total). To build the models,  we implemented cross-validation of 10 iterations using the datasets shown in Table \ref{tabla:datasets}. In each iteration, a grid search was performed to find the best set of hyperparameter values that lead to the most accurate models. After ten iterations, the best model was chosen (among the ten evaluated). In Table \ref{tabla:hyperparameters}, we show the best set of hyperparameter values with which it was possible to train the best model found for each time window duration. 



\begin{table}[h]
  \begin{center}
    \caption{Best Hyperparameters / Training time}
    \label{tabla:hyperparameters}
    \resizebox{8cm}{!}{
    \begin{tabular}{c|c|c|c|c}
      \toprule 
      \multicolumn{5}{c}{\textbf{Random Forest (RF)}} \\
      \midrule
      \textbf{ - } & \textbf{M1: 5 Sec} & \textbf{M2: 20 Sec} & \textbf{M3: 40 Sec} & \textbf{M4: 60 Sec} \\
      \midrule 
            Max\_depth & 6 & 6 & 6 & 6 \\
            N\_estimators & 300 & 500 & 500 & 300  \\
            \hline
            Training time [S] & 332.8 $\pm$ 6 & 304.4 $\pm$ 8.1 & \textbf{264.4} $\pm$ \textbf{20} & 272.6 $\pm$ 20 \\  
      \bottomrule 
      \toprule
      \multicolumn{5}{c}{\textbf{K-Nearest Neighbor (KNN)}} \\
      \midrule
      \textbf{ - } & \textbf{M5: 5 Sec} & \textbf{M6: 20 Sec} & \textbf{M7: 40 Sec} & \textbf{M8: 60 Sec} \\
      \midrule
           \# Neighbors & 3 & 3 & 3 & 3 \\
           \hline
           Training time [S] & 59.8 $\pm$ 2 & 37.7 $\pm$ 1 & 36.3 $\pm$ 3 & \textbf{26.7} $\pm$ \textbf{1} \\  
      \bottomrule
    \end{tabular}}
  \end{center}
\end{table}

In the testing phase, each model was pre-loaded into the classifier component. Then, the WL2 was replicated eight times per each model for the online evaluation. In these experiments, all packets of WL2 were also tagged to compare the tag of each flow with the classifier's classification result (in real-time). In this phase, both the feature set and the flow tag were not stored. Instead, they were sent to the classifier component.

The metrics used in the testing phase to evaluate the detection performance are based on the number of true positive (\textit{TP}), true negative (\textit{TN}), false positive (\textit{FP}) and false negative (\textit{FN}) classes (DDoS and Benign) obtained during the experiment. The metrics are introduced as follows:

\begin{itemize}
    \item \textit{Accuracy} is the number of correctly detected cases in the total flow sample, considering both classes.
    \begin{equation}\label{eq:Acc}
       \textstyle
       \small
       Accuracy = \frac{TN + TP}{TN + TP + FP + FN}
    \end{equation}
    \item \textit{Recall} represents the proportion of DDoS samples correctly detected in relation to the total DDoS flow samples. It is best known as the True Positive Rate (TPR).
    \begin{equation}\label{eq:TPR}
        \textstyle
        \small
        Recall = TPR = \frac{TP}{TP + FN}
    \end{equation}
    \item \textit{Selectivity} represents the proportion of Benign samples correctly detected in relation to the total Benign flow samples; also known as the True Negative Rate (TNR).
    \begin{equation}\label{eq:TNR}
        \textstyle
        \small
        Selectivity = TNR = \frac{TN}{TN + FP}
    \end{equation}
    \item \textit{F1\_Score} can be seen as the weighted average of the precision and recall metrics values.
    \begin{equation}\label{eq:F1}
        \textstyle
        \small
        F1\_score = \frac{2TP}{2TP + FP + FN}
    \end{equation}
\end{itemize}


\subsection{Experimental Results}

Table \ref{tabla:hyperparameters} shows the average time to calculate the best set of hyperparameters and train the model with them. We notice that KNN is the best option when it comes to the performance of the training phase, as it is faster than RF.

\begin{figure}[] 
    \centering
    \includegraphics[scale=0.28]{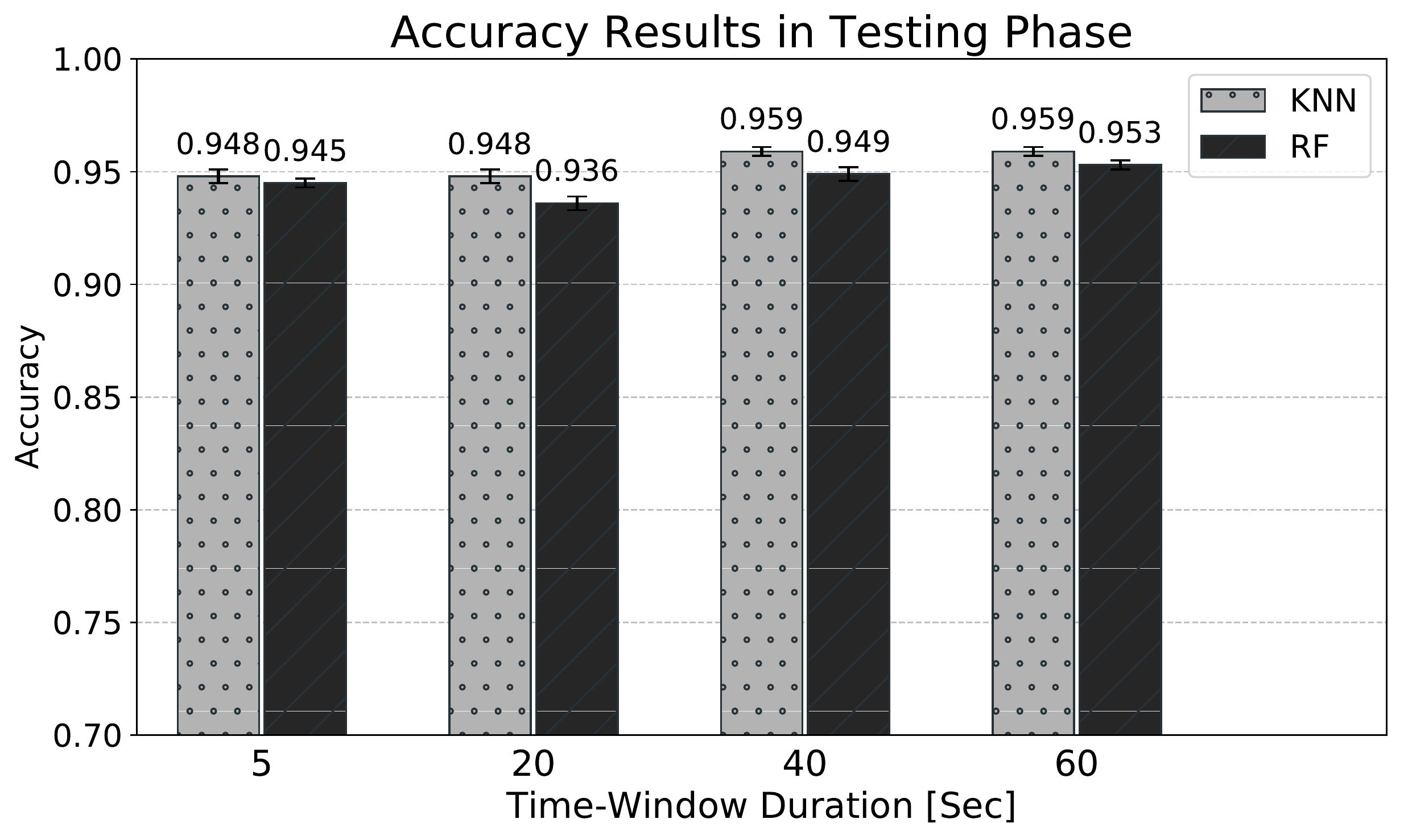}
    \caption{Accuracy of the evaluated models as a function of window duration}
    \label{fig:KNNvsRF}
\end{figure}

Fig. \ref{fig:KNNvsRF} shows the detection accuracy results as a function of the time window duration (in the testing phase). It is possible to observe that the eight models trained lead to an accuracy higher than 93\%, which is within the upper range seen in the literature \cite{IntrusionDS},\cite{atlantic}. Besides, the KNN accuracy results are higher than the RF ones for all time windows. Nevertheless, the difference between models does not exceed 1.2\%.

Increasingly better results are obtained as the time window duration increases, reaching the best values for time windows of 60 seconds.  We highlight here a vital trade-off between accuracy and detection delay. A significant time window duration might lead to an intolerably high latency to detect and mitigate a DDoS attack. We advocate that as the detection rates vary just slightly (1.1\% for KNN and 1.7\% for RF), one should favor lower time window durations.

Table \ref{tabla:metrics} shows the results obtained for the remaining metrics. One can observe how accurate every model is in classifying each class separately (Recall: ``DDoS",  Selectivity: ``Benign"). Also, the F1 score shows the second indicator of accuracy. Note that the main interest of the detection system is to have a classification model very precisely classifying DDoS (low FN) and, at the same time, presenting the least possible number of false alarms (low FP). According to the metrics, a higher Recall value represents a model more accurately detecting DDoS, and a higher Selectivity value represents a model with a lower number of false alarms (see Equations \ref{eq:TPR} and \ref{eq:TNR}). According to that, the M1 model is the most accurate in detecting DDoS attacks, with a Recall value of 98\%. However,  M1 has the lowest selectivity value, which makes this model one of the least reliable, since the false alarm rate would be higher than the other models. This analysis is reflected in the low value of the F1\_Score.

Aiming at a better trade-off among accuracy, false alarms, and time window duration, we consider that M3 and M7, both in the 40-second window, are the best models to use by our detection system. These models have the closest Recall and Selectivity values, which means a more balanced choice concerning the mentioned optimization aspects.


\begin{table}[]
  \begin{center}
    \caption{Comparison of metric results obtained in the testing phase}
    \label{tabla:metrics}
    \resizebox{7.8cm}{!}{
    \begin{tabular}{c|c|c|c}
      \toprule 
      \multicolumn{4}{c}{\textbf{\textbf{Random Forest (RF)}}} \\
      \midrule
      \textbf{T.W. / Metric} & \textbf{Recall} & \textbf{Selectivity} & \textbf{F1 Score} \\
      \midrule 
            M1: 5 seg. & 0.982 $\pm$ 0.001 & 0.920 $\pm$ 0.003 & 0.934 $\pm$ 0.002 \\
            M2: 20 seg. & 0.933 $\pm$ 0.005 & 0.939 $\pm$ 0.003 & 0.933 $\pm$ 0.003 \\
            \textbf{M3: 40 seg.} & \textbf{0.957} $\pm$ \textbf{0.002} & \textbf{0.957} $\pm$ \textbf{0.003} & \textbf{0.952} $\pm$ \textbf{0.003} \\
            M4: 60 seg. & 0.963 $\pm$ 0.003 & 0.941 $\pm$ 0.002 & 0.959 $\pm$ 0.002 \\
      \bottomrule
      \toprule 
      \multicolumn{4}{c}{\textbf{K-Nearest Neighbor (KNN)}} \\
      \midrule
      \textbf{T.W. / Metric} & \textbf{Recall} & \textbf{Selectivity} & \textbf{F1 Score} \\
      \midrule 
            M5: 5 seg. & 0.967 $\pm$ 0.002 & 0.935 $\pm$ 0.003 & 0.937 $\pm$ 0.003 \\
            M6: 20 seg. & 0.951 $\pm$ 0.003 & 0.943 $\pm$ 0.004 & 0.945 $\pm$ 0.003 \\
            \textbf{M7: 40 seg.} & \textbf{0.970} $\pm$ \textbf{0.003} & \textbf{0.967} $\pm$ \textbf{0.003} & \textbf{0.962} $\pm$ \textbf{0.002} \\
            M8: 60 seg. & 0.970 $\pm$ 0.003 & 0.946 $\pm$ 0.003 & 0.964 $\pm$ 0.002 \\
      \bottomrule
    \end{tabular}}
  \end{center}
\end{table}



\section{Conclusion and Future Work}

In this paper, we presented ORACLE, an architecture implemented over an SDN/P4 environment that promotes the cooperation of both the control and data planes to detect network attacks using Machine Learning models. ORACLE leverages data plane programmability to extract flow information at a higher level of granularity. This results in a more straightforward calculation and classification of the feature set in the attack detection process. As proof of concept, we implemented a DDoS detection system based on the computation of intra-flow level features, which cannot be implemented in a traditional SDN/OpenFlow environment. Obtained results show that the detection system has an accuracy rate of up to 96\% while reducing the processing complexity performed by the controller. As future work, we envisage making ORACLE a comprehensive detection system capable of detecting different types of attacks.


\bibliographystyle{IEEEtran.bst}
\bibliography{bibliofile}

\begin{thebibliography}{1}
\providecommand{\url}[1]{#1}
\csname url@samestyle\endcsname
\providecommand{\newblock}{\relax}
\providecommand{\bibinfo}[2]{#2}
\providecommand{\BIBentrySTDinterwordspacing}{\spaceskip=0pt\relax}
\providecommand{\BIBentryALTinterwordstretchfactor}{4}
\providecommand{\BIBentryALTinterwordspacing}{\spaceskip=\fontdimen2\font plus
\BIBentryALTinterwordstretchfactor\fontdimen3\font minus
  \fontdimen4\font\relax}
\providecommand{\BIBforeignlanguage}[2]{{%
\expandafter\ifx\csname l@#1\endcsname\relax
\typeout{** WARNING: IEEEtran.bst: No hyphenation pattern has been}%
\typeout{** loaded for the language `#1'. Using the pattern for}%
\typeout{** the default language instead.}%
\else
\language=\csname l@#1\endcsname
\fi
#2}}
\providecommand{\BIBdecl}{\relax}
\BIBdecl

\bibitem{norton}
\BIBentryALTinterwordspacing
S.~Weisman. (2019) What is a distributed denial of service attack (ddos) and
  what can you do about them? norton. [Online]. Available:
  \url{https://us.norton.com/internetsecurity-emerging-threats-what-is-a-ddos-attack-30sectech-by-norton.html}
\BIBentrySTDinterwordspacing

\bibitem{ChicaJP}
J.~C.~C. Chica~et al., ``Security in sdn: A comprehensive survey,''
  \emph{Journal of Network and Computer Applications}, p. 102595, 2020.

\bibitem{Sultana2019}
N.~Sultana~et al., ``Survey on sdn based network intrusion detection system
  using machine learning approaches,'' \emph{Peer-to-Peer Networking and
  Applications}, vol.~12, no.~2, pp. 493--501, Mar 2019.

\bibitem{p4}
P.~Bosshart~et al., ``P4: Programming protocol-independent packet processors,''
  \emph{SIGCOMM CCR}, vol.~44, no.~3, pp. 87--95, Jul. 2014.

\bibitem{typeflows}
T.~T. Nguyen and G.~Armitage, ``A survey of techniques for internet traffic
  classification using machine learning,'' \emph{IEEE communications surveys \&
  tutorials}, vol.~10, no.~4, pp. 56--76, 2008.

\bibitem{IntrusionDS}
G.~A. Ajaeiya~et al., ``Flow-based intrusion detection system for sdn,'' in
  \emph{IEEE ISCC 2017}.\hskip 1em plus 0.5em minus 0.4em\relax IEEE, 2017, pp.
  787--793.

\bibitem{atlantic}
A.~S. da~Silva~et al., ``Atlantic: A framework for anomaly traffic detection,
  classification, and mitigation in sdn,'' in \emph{IEEE NOMS 2016}.\hskip 1em
  plus 0.5em minus 0.4em\relax IEEE, 2016, pp. 27--35.

\bibitem{yoon2015enabling}
C.~Yoon~et al., ``Enabling security functions with sdn: A feasibility study,''
  \emph{Computer Networks}, vol.~85, pp. 19--35, 2015.

\bibitem{CICIDS2017}
I.~Sharafaldin, A.~H. Lashkari, and A.~A. Ghorbani, ``Toward generating a new
  intrusion detection dataset and intrusion traffic characterization.'' in
  \emph{ICISSP}, 2018, pp. 108--116.

\end{thebibliography}

\end{document}